# Collective excitation of electric dipole on molecular dimer in organic dimer-Mott insulator

K. Itoh[1], H. Itoh[1,2], M. Naka[1], S. Saito[3], I. Hosako[3], N. Yoneyama[2,4],

S. Ishihara[1,2], T. Sasaki[2,5], and S. Iwai*[1,2]

[1] Department of Physics, Tohoku University, Sendai 980-8578, Japan

[2] JST, CREST, Sendai 980-8578, Japan

[3] NICT, Koganei 185-8795, Japan

[4] Interdisciplinary Graduate School of Medicine and Engineering, University of Yamanashi, Kofu, Yamanashi 400-8511, Japan

[5] Institute for Materials Research, Tohoku University, Sendai 980-8577, Japan

* author to whom correspondence should be addressed

**abstract**

The terahertz (THz) response in 10-100 cm$^{-1}$ was investigated in an organic dimer-Mott (DM) insulator $\kappa$-(ET)$_2$Cu$_2$(CN)$_3$ that exhibits a relaxor-like dielectric anomaly. ~30 cm$^{-1}$ band in the optical conductivity was attributable to collective excitation of the fluctuating intra-dimer electric dipoles which are formed by an electron correlation. We succeeded in observing photoinduced enhancement of this ~30 cm$^{-1}$ band, reflecting the growth of the electric dipole cluster in the DM phase. Such optical responses in $\kappa$-(ET)$_2$Cu$_2$(CN)$_3$ reflect instability near the boundary between the DM-ferroelectric charge ordered phases.

In strongly correlated electron systems, the dipole field induced by the Coulomb repulsion sometimes shows ferroelectricity or ferroelectric fluctuation, if the spatial inversion symmetry is broken (1–5). Such a new class of ferroelectric behavior with an electronic origin could exhibit collective response to the external field in ultrafast (~ps or ~1/THz) time scale that enables us to expect applications in Tb/s modulation of ferroelectric memory.

The layered triangular organic dimer Mott (DM) insulator, $\kappa$-(ET)$_2$Cu$_2$(CN)$_3$ [ET; bis(ethylenedithio)-tetrathiafulvalene] is recognized as a spin-liquid system(6–11). Very recently, this compound was also shown to exhibit a relaxor-like dielectric anomaly below 40 K with strong dispersion relation, reflecting the inhomogeneous and fluctuating aspects of the electric dipoles (12–16). In $\kappa$- type ET salts, adjacent ET molecules are dimerized and a charge is localized on each dimer, as shown in Fig. 1(a), if the effective on-site Coulomb repulsion energy $U_{dimer}$ is sufficiently greater than the inter-dimer transfer energy $t$. However, the DM phase is unstable with regard to the ferroelectric charge order (FCO) that is formed by the intra-dimer charge disproportionation for large intermolecular Coulomb energy $V$ or $t$ [Fig. 1(b)](13, 17). Possible origin of the dielectric anomaly in $\kappa$-(ET)$_2$Cu$_2$(CN)$_3$ is the electronic symmetry breaking in the dimer, and the electric dipole cluster state with strong charge fluctuation is considered to be formed in the vicinity of the FCO phase boundary (12-14) (red area in Fig. 1).

In the organic compounds exhibiting FCO (18), the collective excitations, which are attributable to the CDW-like low frequency (100-

MHz) domain wall motions, have been reported (19, 20). On the other hand, collective excitation of electric dipole is expected to be observed at higher frequency (such as THz) region in $\kappa$-(ET)$_2$Cu$_2$(CN)$_3$, because of the dipole is formed by the electronic interaction (12, 13). Furthermore, optical excitation of the DM state competing with FCO enables us to achieve response that is more dramatic than the photoinduced melting of DM state (21).

In this study, we performed steady state terahertz (THz) time domain spectroscopy (TDS) and optical pump-THz probe spectroscopy to detect the optical response of the fluctuating dipole cluster in $\kappa$-(ET)$_2$Cu$_2$(CN)$_3$. In the steady state THz conductivity $\sigma$ spectrum, the broad peak (E//c) at ~30 cm$^{-1}$ (~1 THz) was attributable to the collective excitation of the intra-dimer electric dipole. Optical pump-THz probe measurement demonstrated that this THz response was increased by photoexcitation of the DM state, showing the photoinduced growth of the electric dipole cluster state.

Single crystals of $\kappa$-(ET)$_2$Cu$_2$(CN)$_3$ (average size: 1 × 1 × 0.5 mm$^3$) were prepared using a previously reported procedure (11). Steady state conductivity spectra were directly obtained from waveform in time domain detected by TDS transmission measurement. Near-infrared (NIR) pump (0.89 eV)-THz-probe TDS (10–100 cm$^{-1}$) spectroscopy was performed using a 1-kHz Ti:Al$_2$O$_3$ regenerative amplifier system (Legend Elite USX; Coherent). The NIR pump pulse was generated in an optical parametric amplifier, and the THz probe pulse was emitted in a ZnTe crystal; they were focused on a single crystal of $\kappa$-(ET)$_2$Cu$_2$(CN)$_3$ in a 1.5-mm-diameter region excited by the pump beam. The THz probe was detected by electro-optic (EO) sampling. The time resolution was ~ 1 ps.

Figures 2(a) and 2(b) show the steady state optical conductivity σ(ω) spectra at 10 K for E//b [Fig. 2 (a)] and E//c [Fig. 2(b)]. Several phonon peaks were observed at 23, 39, 41, 48, 56, 69, 84, 87, 97, 92 cm$^{-1}$ for E//b and at 41, 51, 59, 66, 69, 73, 83 cm$^{-1}$ for E//c. The inset of Fig. 2(b) shows the changes of the peak energies as $E_{shift} = E_p(T)/E_p(80 K)$ ($E_p(T)$ represents the peak energy indicated by the blue arrows at temperature $T$). Red (51 cm$^{-1}$) and blue (41, 83 cm$^{-1}$) shifts of the peaks with reducing temperature are attributable to the anisotropic thermal expansion and shrinkage of the lattice (11). Small (< 2%) and continuous changes of $E_{shift}$ reflect the absence of the marked structural transition. In addition to these phonon peaks, we also detected the broad spectrum at ~30 cm$^{-1}$ for E//c [dashed circle in Fig. 2(b)]. We will refer to this band as "30 cm$^{-1}$ band". Fig. 2(c) shows the temperature dependence of the 30 cm$^{-1}$ band, indicating the marked increase of this band at low temperature. A prominent spectral dip at the center of this band is attributable to the Fano interference between the broad electronic and sharp phonon excitations as discussed below. Here, the spectral shape of the 30 cm$^{-1}$ band was analyzed by using the analytical Fano's expression for a broad electronic transition that contains underlying dispersion (22)

$$\sigma(\omega) = -\frac{A}{\pi} \text{Im} \frac{S_{dp}(\omega)}{1 - \gamma'^2 S_{dp}(\omega) S_{ph}(\omega)}$$

where $S_{ph}(\omega)$ and $S_{dp}(\omega)$ represent spectral functions (Lorentz functions with the peak energies of $E_{ph}$ and $E_{dp}$) for an optically forbidden phonon excitation and the collective dipole excitation, respectively. The coupling constant between both excitations is represented by $\gamma'$. As shown by the solid curves in Fig. 2(c), the observed spectra can be reproduced by the calculated curves. Fig. 2 (d) indicates *A,* a coefficient related to the fraction of the electronic state, which is shown to grow markedly at low temperature. Then, as shown in the inset, A saturates at 5.5 K where the characteristic temperature of this dipole anomaly has been observed in various measurements (11, 12). The dielectric relaxation with large dispersion observed below ~40 K (12), that is widely seen in random systems such as relaxor ferroelectrics (23) indicates the emergence of the dipole cluster in the DM phase. Considering that, the marked increase in the 30 cm$^{-1}$ band at low temperature is associated with the growth of the dipole cluster.

We examined theoretically collective charge excitations in an organic dimer Mott insulator (13, 24). Since a size of the dipole cluster is expected to be much larger than a coherent length of the collective mode, as shown later, here we present a theoretical description in a homogeneous system. The extended Hubbard model and the spin-less *V-t* model with the internal-degree of freedom of dimers are analyzed by the exact diagonalization method combined with the mean-field approximation. Low energy parts of the excitations were examined, in more detail, on the effective pseudo-spin (PS) model, where electronic structure in a dimer molecule is described by the PS operator **Q** with an amplitude of 1/2; eigen states of $Q^z$ and $Q^x$ correspond to charge depolarized and polarized states, respectively (13, 24), where the electronic charge distributes on a dimer with and without the space-inversion symmetry, respectively. There is a charge excitation from the one-state to another. These naively imply as charge redistribution inside a dimer. In a crystal where dimers are arranged at lattice sites, the intra-dimer charge excitation can propagate due to the inter-dimer electronic interactions. Since the charge redistribution occurs dimer by dimer, we recognize it a collective excitation for the electronic dipole, showing dispersion relation.

Excitation energies of the collective excitations are of the order of the inter-dimer electron transfer (the inter-site Coulomb interaction) in the DM (FCO) phase far from the phase boundary. Calculations show that the collective mode softens, and the excitation energy at the Brillouin zone center becomes zero when the system approaches the boundary. This is because the phase transition between the two phases is of the second order. Our theory predicts that this soft collective mode is detected by the optical conductivity spectra, when the electric field is parallel to the spontaneous electric polarization predicted theoretically (//c axis) as shown in Fig. 1(b), since the collective excitation corresponds to a change in the electric dipole moment. These characteristics provide evidence that the observed 1THz band is attributed to the collective charge excitation mode.

The spectral analysis using above Fano's analytical expression provides more detailed information. Fig. 2(e) shows that $E_{\mathrm{dp}}$ is almost independent of temperature, although the small (~2 cm$^{-1}$) red shift at low temperature might reflect the softening. The reason of such no or small softening remains unclear. One possible interpretation is that the coherence length (CL) of the collective excitation is much smaller than the size of the dipole cluster. The small CL is also indicated by the fact that the resonant frequency of the collective excitation (~30 cm$^{-1}$) is much higher than the detection frequencies of the dielectric anomaly (0.5- 500 kHz ) (12). On the other hand, $\gamma$ increases with decreasing temperature (80–40 K), but it saturates at low temperature, as shown in Fig. 2 (f). Therefore, the electron-phonon interaction within the CL does not change at low temperature, although the continuous increase of A suggests that dipole clusters continue to grow.

We can notice the spectral broadening at low temperature reflecting the charge fluctuation and/or the inhomogeneity in the previous IR/Raman/NMR studies (25-28), although the clear splitting of the peaks showing the static and homogeneous FCO were not observed. Our interpretation using "fluctuating intra-dimer dipole" picture does not conflict with those previous studies.

In the DM-FCO competing state, instability of the DM phase, leading to the photoinduced DM-FCO conversion, is expected to be induced by optical excitations (21). Figure 3(a) shows the photoinduced changes of optical conductivity $\Delta\sigma$ at $t_d$ (delay time between the pump and THz probe pulses) =0.1 ps (red curve in Fig. 3(a)) after the excitation at 0.89 eV, corresponding to the high energy side of the intermolecular charge transfer (CT) transition including the components of both intra-dimer and inter-dimer excitations (24) (excitation intensity $I_{ex}$=0.01 mJ/cm², 1 photon/ 1000 molecules). These excitations can induce the instability of the DM phase (21). Polarizations of the excitation and probe electric field (E$_{pu}$, E$_{pr}$) are (//c, //c) for red solid curve, (//c, //b) for blue solid curve, and (//b, //c) for black dashed curve, respectively.

The spectral features of $\Delta\sigma$ at $t_d$ =0.1 ps, showing peaks at 29 and 34 cm$^{-1}$ for both excitation polarizations parallel to //c and //b are quite analogous to those of the spectral difference between 6 K and 10 K ($\sigma$(6 K)- $\sigma$(10 K) for //c axis; red curve in Fig. 3(c)). That reflects the increase in the 30 cm$^{-1}$ band, indicating the photoinduced change from the DM state to the intra-dimer dipole state, i.e., the growth of the dipole cluster in the DM phase, if we assume that the 30 cm$^{-1}$ band is attributable to the collective excitation of the intra-dimer dipole. It is noteworthy that such characteristic increase of $\Delta\sigma$ peaked at 29 and 34 cm$^{-1}$ is detected only for $E_{pr}$//c, showing that the photoinduced dipole clusters are polarized along //c. The broad backgrounds of $\Delta\sigma$ for both $E_{pr}$ //b and $E_{pr}$ //c are attributable to the Drude-like excited carrier. The analogous results for both excitation polarizations $E_{pu}$//c (red solid curve in Fig. 3(a)) and $E_{pu}$//b (black dashed curve in Fig. 3(a)) are attributable to the fact that the inter-molecular CT transition occurs similarly for both excitation polarizations.

Such photoinduced growth of the intra-dimer dipole cluster in the DM phase is completely different from the melting of the DM or the FCO phases near the phase boundary with the metallic state which have been reported (21, 29). Figures 3(d) and 3(e) show that the $\Delta\sigma$ ($t_d$=0.1 ps) increases markedly below 20 K. The characteristic increase in $\Delta\sigma$ at low temperatures shows that the instability of the DM phase to the FCO state increases at low temperature as shown in Fig. 1.

Figure 4(a) shows the time evolution of $\Delta\sigma$ observed at 30 cm$^{-1}$, which can be reproduced using the equation:

$$\Delta\sigma(t) = \int_{-\infty}^{t} \left[ Ae^{(-\tau/\tau_{decay})} + Be^{(-\tau/\tau_{damp})} \cos(\omega\tau - \varphi) \right] g(t-\tau) d\tau, \quad g(t) = \frac{1}{\sqrt{\pi}} e^{-(4\ln 2/F^2)t^2}$$

where the time constant of the decay, $\tau_{decay} = 2.6\,ps$, the oscillating period, $2\pi/\omega = 4.6\,ps$, the oscillating damping time $\tau_{damp} = 8\,ps$, the initial phase of cosine function, $\varphi = 0$ and the FWHM of the response function, $g(t)$ $F = 1\,ps$, the Coefficients $A = 0.017$ and $B = 0.014$.

The fast (2.6 ps) recovery of the photoinduced state shows the absence of the large structural difference between the DM and intra-dimer dipole state. The oscillating energy $\hbar\omega = 7.3$ cm$^{-1}$ does not correspond to that of the inter-molecular optical phonons (30-100 cm$^{-1}$) which are observed as the coherent phonons in the mid-infrared pump-probe measurement (30). A possible candidate for the origin of this oscillation is the optical mode which is softened as a result of the coupling with the electronic and/or the acoustic phonon modes. That might be related to the domain wall motion of the photoinduced dipole cluster. However, the time scale of this oscillation (4.6 ps) is much faster than those of the CDW-like DW motions observed in other organic conductor (19, 20). The DW motion observed here is different from those slower DW motions and is considered to be the more microscopic scale.

In summary, a broad peak (E//c) was observed at ~30 cm$^{-1}$ reflecting the collective excitation of the intra-dimer electric dipole in organic dimer insulator κ-(ET)$_2$Cu$_2$(CN)$_3$. Optical pump-THz probe measurement showed that this THz response was increased by the photoexcitation, demonstrating the photoinduced growth of the electric dipole cluster, as a result of the photoinduced collapse of the DM phase competing with the FCO phase.

Figure caption;

FIG. 1 Schematic illustration of the theoretically predicted phase diagram of a DM insulator κ-ET salt (13, 14, 16, 17). Molecular arrangement and charge distribution of DM phase (a) and FCO phase (b) are shown. Red arrow shows the direction of the macroscopic polarization P in FCO phase.

FIG. 2 (a), (b) Steady state optical conductivity σ(ω) spectra at 10 K for E//b (a) and E//c (b). Inset shows temperature dependences of the peak energy changes ($E_{shift}$) for the peaks indicated by the blue arrows. (c) Temperature dependence of the THz response (20–38 cm$^{-1}$) observed for E//c. (d), (e), (f) Temperature dependence of $A$ (d), $E_{ph}$, $E_{dp}$ (e) and γ (f) (see text).

FIG.3 (a) Photoinduced changes of optical conductivity Δσ at $t_d$ =0.1 ps (red curve for ($E_{pu}$, $E_{pr}$)= (//c, //c), black dashed curve for (//b, //c) and blue curve for (//c, //b)). (b) Steady state σ spectra for //c (c) Spectral differences σ(6 K)- σ(10 K) (red curve) and σ(20 K)- σ(10 K) (blue curve) (d) Δσ spectrum at various temperatures (e) Temperature dependences of Δσ observed at 34 and 29 cm$^{-1}$.

FIG. 4 Time evolution of Δσ observed at 29 cm$^{-1}$ (open circles) and fitting curve (red curve). The solid curve shows the oscillating component.

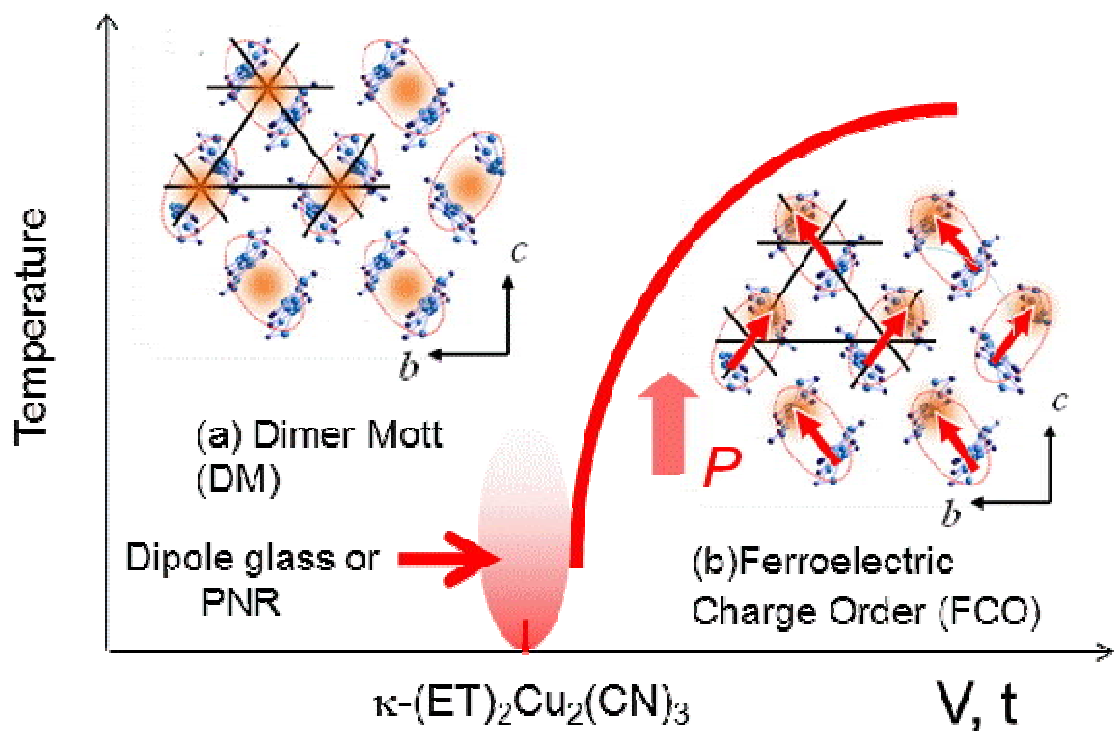

Fig. 1 Itoh et al

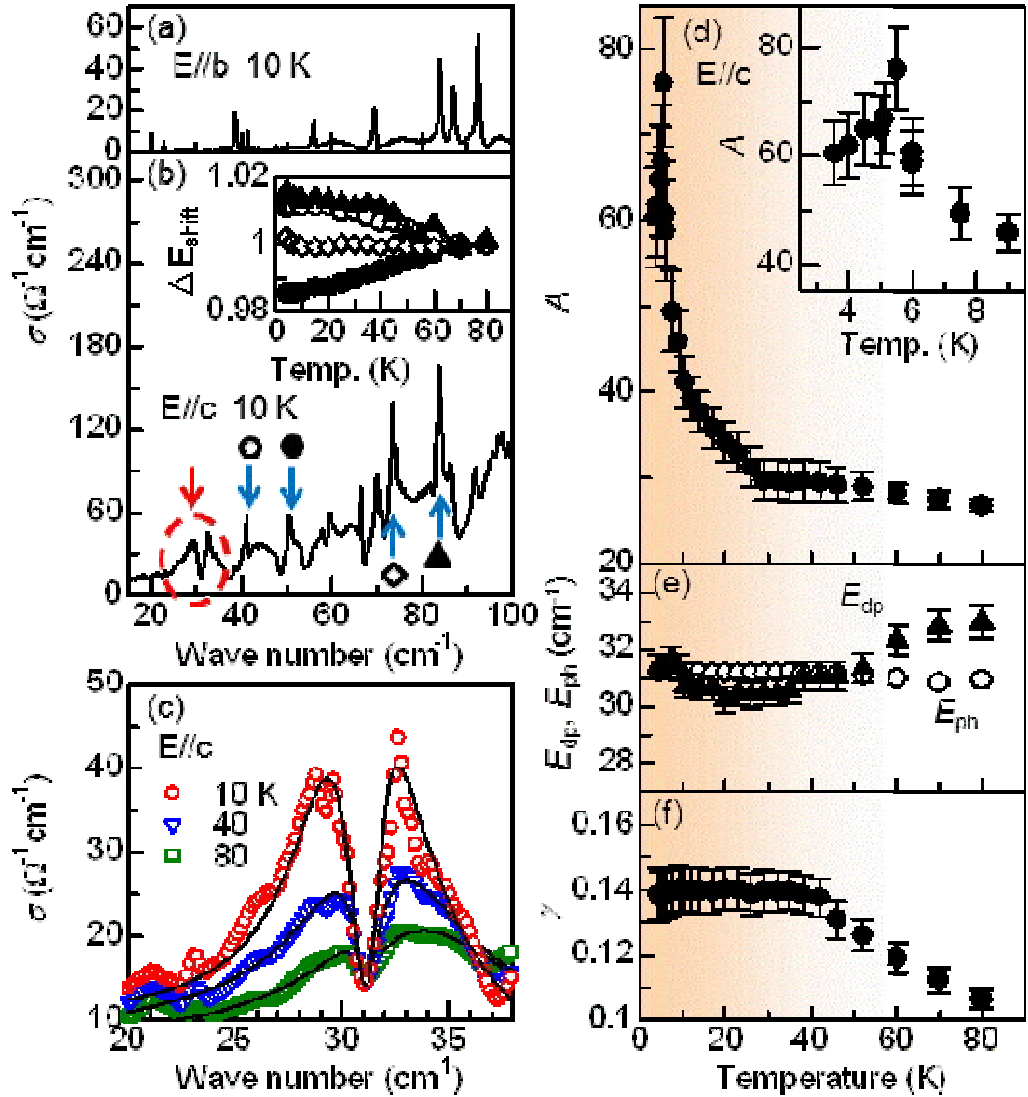

Fig. 2 Itoh et al

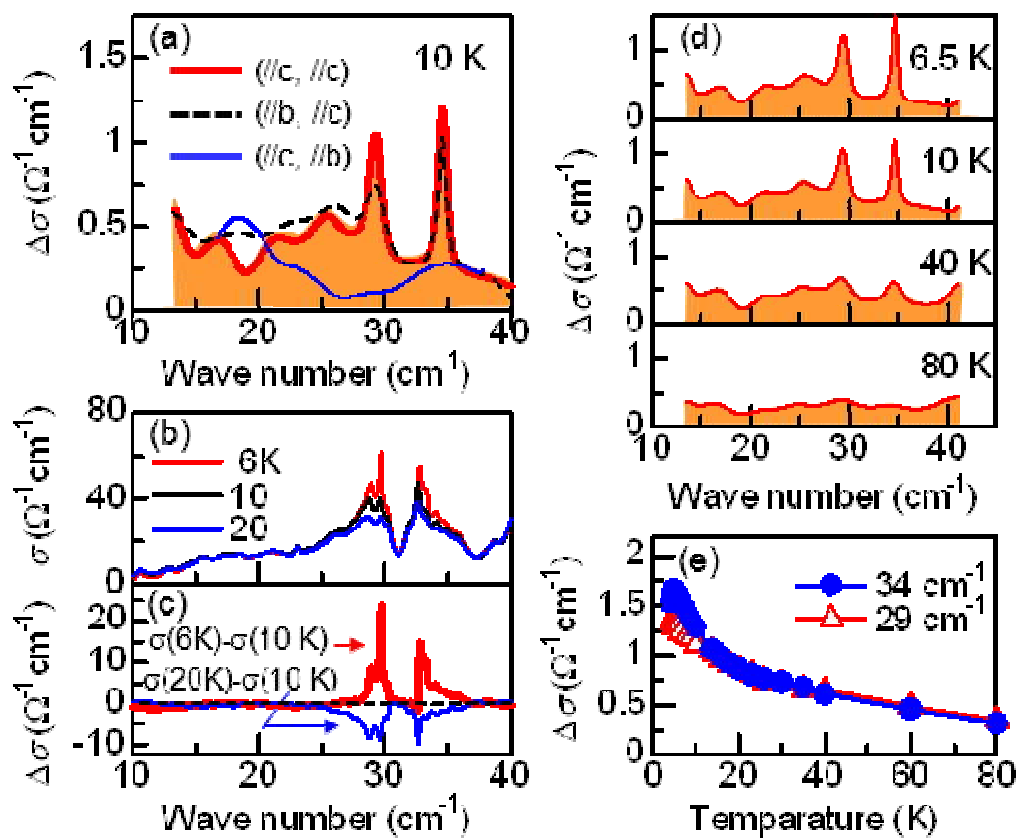

Fig. 3 Itoh et al

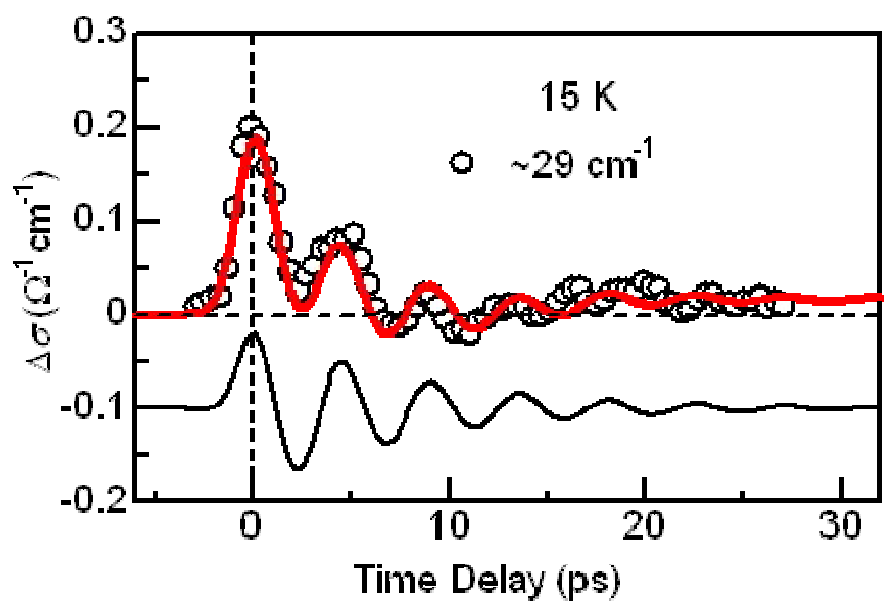

Fig. 4  Itoh et al